\documentclass[conf, 10pt, a4paper]{IEEEtran}

\usepackage{color}
\usepackage{graphicx}
\usepackage{amsmath, amsfonts, amssymb, psfrag, epsfig, epstopdf, bbm}
\usepackage{footnote}
\newcommand\scalemath[2]{\scalebox{#1}{\mbox{\ensuremath{\displaystyle #2}}}}
\newtheorem{proposition}{{Proposition}}
\newtheorem{definition}{{Definition}}
\newtheorem{theorem}{{Theorem}}

\newtheorem{example}{{Example}}

\newtheorem{remark}{{Remark}}

\DeclareMathAlphabet{\mathpzc}{OT1}{pzc}{m}{it}
\DeclareMathOperator*{\argmax}{arg\, max}

\allowdisplaybreaks

\newdimen\slantmathcorr
\def\oversl#1{%assuming that mathslant=0.25
\setbox0=\hbox{$#1$}
\slantmathcorr=\wd0
\hskip 0.2\slantmathcorr \overline{\hbox to 0.8\wd0{%
\vphantom{\hbox{$#1$}}}}
\hskip-\wd0\hbox{$#1$}
}

\def\undersl#1{%assuming that mathslant=0.25
\setbox0=\hbox{$#1$}
\slantmathcorr=\wd0
\underline{\hbox to 0.8\wd0{%
\vphantom{\hbox{$#1$}}}}
\hskip-0.8\wd0\hbox{$#1$}
}


\newcounter{parentnumber}

\begin{document}

\title{Sufficient Conditions for the Tightness of Shannon's Capacity Bounds for Two-Way Channels}
%Sufficient Conditions for Determining the Capacity Region of Two-Way Channels

\author{
\IEEEauthorblockN{Jian-Jia Weng\IEEEauthorrefmark{2}, Lin Song\IEEEauthorrefmark{3}, Fady Alajaji\IEEEauthorrefmark{2}, and Tam\'as Linder\IEEEauthorrefmark{2}}
\thanks{% 
    \IEEEauthorrefmark{2}The authors are with the Department of Mathematics and Statistics, Queen's University, Kingston, ON K7L 3N6, Canada (Emails: jian-jia.weng@queensu.ca, \{fady, linder\}@mast.queensu.ca).}
\thanks{%
    \IEEEauthorrefmark{3} L.~Song was with the Department of Mathematics and Statistics, Queen's University, Kingston, ON K7L 3N6, Canada. She is now with Contextere Ltd., Ottawa, ON K1Y 2C5, Canada  (Email: lin@contextere.com).}
\thanks{
This work was supported in part by NSERC of Canada.}
}

\maketitle
\begin{abstract}
New sufficient conditions for determining in closed form the capacity region of point-to-point memoryless two-way channels (TWCs) are derived. 
The proposed conditions not only relax Shannon's condition which can identify only TWCs with a certain symmetry property but also generalize other existing results. 
Examples are given to demonstrate the advantages of the proposed conditions. 
\end{abstract}


\section{Introduction}\label{sec:introduction}
Finding the capacity region of point-to-point discrete memoryless two-way channels (TWCs) in single-letter form is a long-standing open problem. 
The difficulty lies in the causality of transmission, since the senders are allowed to generate channel inputs by adapting to previously received channel outputs.
In \cite{Shannon:1961}, Shannon gave an (uncomputable) multi-letter expression for the capacity region. 
Another multi-letter expression, using directed information \cite{Massey:1990}, was given in \cite{Krama:1998}.
The capacity region of TWCs is known only for some special channels such as TWCs with additive white Gaussian noise \cite{Han:1984}, determinisitc TWCs \cite{Devroye:2014}, TWCs with discrete additive noise \cite{Song:2016}, and injective semi-deterministic TWCs \cite{Chaaban:2017}. 
Thus, Shannon's inner and outer bounds \cite{Shannon:1961} still play an important role in characterizing the capacity region. 

In the literature, Shannon's symmetry condition \cite{Shannon:1961} and a condition established by Chaaban, Varshney, and Alouini (CVA) \cite{Chaaban:2017} are two known sufficient conditions under which Shannon's inner and outer bounds coincide, thus directly characterizing the capacity region. 
Shannon's condition focuses on a certain symmetry structure for the channel transition probabilities, while the CVA condition focuses on the existence of independent inputs which achieve Shannon's outer bound. 
Although the two conditions can be used to determine the capacity region of a large class of TWCs, it is of interest to establish new conditions for wider families of channels. 

In this paper, four sufficient conditions guaranteeing that Shannon's inner and outer bounds coincide are derived. 
Similar to the CVA condition, our conditions identify independent inputs which achieve Shannon's outer bound based on the approach that a TWC can be viewed as two one-way channels with state.
Two of the derived results are shown to be substantial generalizations of the Shannon and CVA conditions. 
Moreover, our simplest condition can be easily verified by observing the channel marginal distributions. 

The rest of this paper is organized as follows. 
In Section~II, the system model and prior results are reviewed.  
New conditions for finding the capacity region are provided in Section~III.
A discussion of the connections between the new conditions and prior results is given in Section~IV along with illustrative examples. 
Concluding remarks are given in Section~V. 

\begin{figure}[!tb]
\begin{centering}
\vspace{-0.2cm}
\includegraphics[scale=0.425, draft=false]{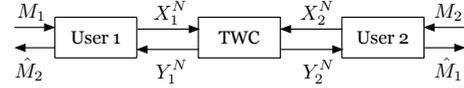}
\caption{Block diagram of two-way transmission.\label{fig:TWModel}}
\vspace{-0.45cm}
\end{centering}
\end{figure}
%\vspace{-0.27cm}

\section{Preliminaries}\label{sec:p2p}
In a two-way communication system as shown in Fig. \ref{fig:TWModel}, two users want to exchange their own messages $M_1$ and $M_2$ via $N$ uses of a TWC. 
Here, the messages $M_1$ and $M_2$ are assumed to be mutually independent and uniformly distributed on $\mathcal{M}_1 \triangleq \{ 1,2,..., 2^{NR_1} \} $ and $\mathcal{M}_2 \triangleq \{ 1,2,...,2^{NR_2} \}$, respectively, where $NR_1$ and $NR_2$ are non-negative integers. 
For $j=1, 2$, let $\mathcal{X}_j$ and $\mathcal{Y}_j$ respectively denote the finite channel input and output alphabets for user $j$. 
The joint distribution of the inputs and outputs of a memoryless TWC is governed by the channel transition probability $P_{Y_1, Y_2|X_1, X_2}$.  
A channel code for a TWC is defined as follows. 

\begin{definition}
An $(N, R_1, R_2)$ code for a TWC consists of two message sets $\mathcal{M}_1=\{1, 2, \dots, 2^{NR_1}\}$ and $\mathcal{M}_2=\{1, 2, \dots, 2^{NR_2}\}$, two sequences of encoding functions $f_1^N\triangleq (f_{1,1}, f_{1,2}, \dots, f_{1,N})$ and $f_2^N\triangleq (f_{2,1}, f_{2,2}, \dots, f_{2,N})$, with $f_{1,1}:  \mathcal{M}_1 \to \mathcal{X}_1$, $f_{1,n}:  \mathcal{M}_1 \times  \mathcal{Y}_1^{n-1} \to \mathcal{X}_1$, $f_{2, 1}: \mathcal{M}_2 \to \mathcal{X}_2$, and $\mathcal{M}_2 \times  \mathcal{Y}_2^{n-1} \to \mathcal{X}_2$
%\[
%\begin{array}{ll}
%f_{1,1}:  \mathcal{M}_1 \to \mathcal{X}_1,& f_{1,n}:  \mathcal{M}_1 \times  \mathcal{Y}_1^{n-1} \to \mathcal{X}_1\\
%f_{2,1}:  \mathcal{M}_2 \to \mathcal{X}_2,& f_{2,n}:  \mathcal{M}_2 \times  \mathcal{Y}_2^{n-1} \to \mathcal{X}_2
%\end{array}
%\]
for $n=2, 3, \dots, N$, and two decoding functions $g_1: \mathcal{M}_1 \times  \mathcal{Y}_1^{N} \to \mathcal{M}_2$ and $g_2: \mathcal{M}_2 \times  \mathcal{Y}_2^{N} \to \mathcal{M}_1$. 
\end{definition}

When messages $M_1$ and $M_2$ are encoded, the channel inputs at time $n=1$ are only functions of the messages, i.e., $X_{j, 1} = f_{j, 1}(M_j)$ for $j=1, 2$, but all the other channel inputs are generated by also adapting to the previous channel outputs $Y_j^{n-1}\triangleq(Y_{j, 1}, Y_{j, 2}, \dots, Y_{j, n-1})$ via $X_{j, n} = f_{j, n}(M_j, Y_j^{n-1})$ for $j=1,2$ and $n=2, 3, \dots, N$.
After receiving $N$ channel outputs, user $j$ reconstructs $M_i$ as $\hat{M}_i =g_j(M_j, Y_j^N)$ for $i, j=1, 2$ with $i\neq j$, and the probability of decoding error is defined as $P^{(N)}_{e}(f_1^N, f_2^N, g_1, g_2)=\text{Pr}\{\hat{M}_1 \neq M_1\ \text{or}\ \hat{M}_2 \neq M_2\}.$
Based on this performance index, we define achievable rate pairs and the capacity region. 

\begin{definition}
A rate pair $(R_1,R_2)$ is said to be achievable if there exists a sequence of $(N, R_1, R_2)$ codes such that $\lim_{N \to \infty} P^{(N)}_{e}=0.$
The capacity region $\mathcal{C}$ of a TWC is the closure of the convex hull of all achievable rate pairs. 
\end{definition}

To date, a computable single-letter expression for the capacity region of general memoryless TWCs has not been found. 
In \cite{Shannon:1961}, Shannon established inner and outer bounds for the capacity region. 
Let $\mathcal{R}(P_{X_1,X_2},P_{Y_1,Y_2|X_1,X_2})$ denote the set of rate pairs $(R_1,R_2)$ with $R_1 \le I(X_1;Y_2|X_2)$ and $R_2 \le I(X_2;Y_1|X_1)$, 
 where the joint distribution of all random variables is given by $P_{X_1,X_2}P_{Y_1,Y_2|X_1,X_2}$. 
 Then, the capacity region of a discrete memoryless TWC with transition probability $P_{Y_1,Y_2|X_1,X_2}$ is inner bounded by \cite{Shannon:1961}
 \begin{align*}
 \scalemath{0.97}{\mathcal{C}_I (P_{Y_1,Y_2|X_1,X_2}) \triangleq \scalemath{0.9}{\overline{\text{co}} \left( \bigcup_{P_{X_1}P_{X_2}} \mathcal{R}(P_{X_1}P_{X_2}, P_{Y_1,Y_2|X_1,X_2}) \right)}},
 \end{align*}
 and outer bounded by
 \begin{align*}
\scalemath{0.97}{\mathcal{C}_O (P_{Y_1,Y_2|X_1,X_2}) \triangleq \scalemath{0.9}{\overline{\text{co}} \left( \bigcup_{P_{X_1,X_2}} \mathcal{R}(P_{X_1,X_2}, P_{Y_1,Y_2|X_1,X_2}) \right)}},
  \end{align*}
where $\overline{\text{co}}$ denotes taking the closure of the convex hull.
In general, $\mathcal{C}_I$ and $\mathcal{C}_O$ are not matched, but if they coincide, then the exact capacity region is obtained by independent inputs. 
We note there exist other improved bounds for TWCs \cite{Han:1984}, \cite{Schalkwijk:1982}-\cite{Hekstra:1989}. However, those bounds are either restricted to the binary multiplier TWC \cite{Schalkwijk:1982}, \cite{Schalkwijk:1983} or expressed with auxiliary random variables \cite{Han:1984, Zhang:1986, Hekstra:1989}, which do not fit the needs of our approach. 
It is worth mentioning that for single-output TWCs, i.e., when $Y_1=Y_2$, a condition for $\mathcal{C}_I=\mathcal{C}_O$ is given in \cite{Hekstra:1989}.

We next review the Shannon \cite{Shannon:1961} and CVA \cite{Chaaban:2017} conditions that imply the equality of $\mathcal{C}_I$ and $\mathcal{C}_O$. 
For a finite set $\mathcal{A}$, let $\pi^\mathcal{A}: \mathcal{A} \to \mathcal{A}$ denote a permutation (bijection), and for any two symbols $a'$ and $a''$ in $\mathcal{A}$, let $\tau^{\mathcal{A}}_{a',a''}: \mathcal{A} \to \mathcal{A}$ denote the transposition which swaps $a'$ and $a''$ in $\mathcal{A}$, but leaves the other symbols unaffected. 
Moreover, let $P_{X, Z, Y}=P_{X}P_{Z|X}P_{Y|X, Z}$ denote a probability distribution defined on finite sets $\mathcal{X}$, $\mathcal{Y}$, and $\mathcal{Z}$. 
We define two functionals for conditional entropies:
\begin{IEEEeqnarray}{l}
\scalemath{0.9}{\mathcal{H}(P_{X, Z}, P_{Y|X, Z})}\triangleq \scalemath{0.85}{\sum\limits_{x, z, y}P_{X, Z}(x, z)P_{Y|X, Z}(y|x, z) \log\frac{1}{P_{Y|X, Z}(y|x, z)}}\nonumber\label{H1}
\end{IEEEeqnarray}
and 
\begin{IEEEeqnarray}{l}
\scalemath{0.9}{\bar{\mathcal{H}}(P_{X}, P_{Z|X}, P_{Y|X, Z})}\triangleq\scalemath{0.85}{\sum\limits_{x, y}P_{X}(x)P_{Y|X}(y|x)\log\frac{1}{P_{Y|X}(y|x)}},\nonumber\label{H2}
\end{IEEEeqnarray}
where $P_{Y|X}(y|x)=\sum_{z}P_{Y|X, Z}(y|x, z)P_{Z|X}(z|x)$. 
In particular, if $P_{X, Z}=P_{X}P_{Z}$, we let $P_Z(z)=\sum_x P_{X, Z}(x, z)$ and define  
\begin{IEEEeqnarray}{l}
\scalemath{0.9}{\mathcal{\bar{H}}_{\perp}(P_{X}, P_{Z}, P_{Y|X, Z})}\triangleq\scalemath{0.85}{\sum\limits_{x, y}P_{X}(x)Q_{Y|X}(y|x)\log\frac{1}{Q_{Y|X}(y|x)}},\nonumber\label{H3}
\end{IEEEeqnarray}
where $Q_{Y|X}(y|x)=\sum_{z}P_{Y|X, Z}(y|x, z)P_{Z}(z)$. 

Note that, given any $P_{X_1, X_2}=P_{X_2}P_{X_1|X_2}=P_{X_1}P_{X_2|X_1}$, we have $H(Y_j|X_1, X_2)=\mathcal{H}(P_{X_1, X_2}, P_{Y_j|X_1, X_2})$, $H(Y_1|X_1)=\allowbreak\bar{\mathcal{H}}(P_{X_1}, P_{X_2|X_1}, P_{Y_1|X_1, X_2})$, and $H(Y_2|X_2)=\allowbreak\bar{\mathcal{H}}(P_{X_2}, P_{X_1|X_2}, P_{Y_2|X_1, X_2})$, where $P_{Y_j|X_1, X_2}$ is a marginal of the channel probability $P_{Y_1, Y_2|X_1, X_2}$ and $j=1, 2$. 
Furthermore, for any $P_{X_1, X_2}=P_{X_1}P_{X_2}$, we have $H(Y_1|X_1)=\allowbreak\mathcal{\bar{H}}_{\perp}(P_{X_1}, P_{X_2}, P_{Y_1|X_1, X_2})$ and $H(Y_2|X_2)=\allowbreak\mathcal{\bar{H}}_{\perp}(P_{X_2}, P_{X_1}, P_{Y_2|X_1, X_2})$.
Finally, let $\mathcal{P}(\mathcal{X}_j)$ denote the set of all probability distributions on $\mathcal{X}_j$ for $j=1, 2$. 

\begin{proposition}[Shannon's Symmetry Condition \cite{Shannon:1961}]\label{thm:SC}
For a memoryless TWC with transition probability $P_{Y_1,Y_2|X_1,X_2}$, we have $\mathcal{C}=\mathcal{C}_I=\mathcal{C}_O$ if for any pair of distinct input symbols $x'_1, \allowbreak x''_1\in\mathcal{X}_1$, there exists a pair of permutations 
$(\pi^{\mathcal{Y}_1}[x'_1,x''_1], \pi^{\mathcal{Y}_2}[x'_1,x''_1])$ on $\mathcal{Y}_1$ and $\mathcal{Y}_2$, respectively, (which depend on $x'_1$ and $x''_1$) such that for all $x_1$, $x_2$, $y_1$, $y_2$, 
\vspace{-0.1cm}
\begin{IEEEeqnarray}{l}
\scalemath{0.94}{P_{Y_1,Y_2|X_1,X_2}(y_1,y_2|x_1,x_2)=}\nonumber\\
\ \ \scalemath{0.9}{P_{Y_1,Y_2|X_1,X_2}(\pi^{\mathcal{Y}_1}[x'_1,x''_1](y_1),\pi^{\mathcal{Y}_2}[x'_1,x''_1](y_2)|\tau^{\mathcal{X}_1}_{x'_1,x''_1} (x_1),x_2).}\IEEEeqnarraynumspace\label{SCcond}
\end{IEEEeqnarray}
%for all $x_1\in\mathcal{X}_1$, $x_2\in\mathcal{X}_2$, $y_1\in\mathcal{Y}_1$, and $y_2\in\mathcal{Y}_2$, then the regions $\mathcal{C}_I$ and $\mathcal{C}_O$ coincide. 
\end{proposition}

\begin{proposition}[CVA Condition \cite{Chaaban:2017}]\label{thm:CVA}
For a memoryless TWC with transition probability $P_{Y_1,Y_2|X_1,X_2}$, we have $\mathcal{C}=\mathcal{C}_I=\mathcal{C}_O$ if for any $P_{X_1,X_2} = P_{X_2}P_{X_1|X_2}=P_{X_1}P_{X_2|X_1}$, $\mathcal{H}(P_{X_2}\tilde{P}_{X_1|X_2}, P_{Y_j|X_1, X_2})$ does not depend on $\tilde{P}_{X_1|X_2}$ for given $P_{X_2}$ and there exists $\tilde{P}_{X_1}\in\mathcal{P}(\mathcal{X}_1)$ such that ${\mathcal{\bar{H}_{\perp}}}(\tilde{P}_{X_1}, P_{X_2}, P_{Y_1|X_1, X_2})\ge \bar{\mathcal{H}}(P_{X_1}, P_{X_2|X_1}, P_{Y_1|X_1, X_2})$ and ${\mathcal{\bar{H}_{\perp}}}(P_{X_2}, \tilde{P}_{X_1}, P_{Y_2|X_1, X_2})\ge \bar{\mathcal{H}}(P_{X_2}, P_{X_1|X_2}, P_{Y_2|X_1, X_2})$. 
\end{proposition}
%\vspace{+0.1cm}

We remark that Proposition~\ref{thm:SC} describes a channel symmetry property with respect to the channel input of user~1, but an analogous condition can be obtained by exchanging the roles of users~1 and~2. Also, the invariance of $\mathcal{H}(P_{X_2}\tilde{P}_{X_1|X_2}, P_{Y_j|X_1, X_2})$ in Proposition~2 in fact imposes a certain symmetry constraint on the channel marginal distribution $P_{Y_j|X_1, X_2}$.
In the literature, a TWC with independent $q$-ary additive noise \cite{Song:2016} is an example that satisfies both the Shannon and CVA conditions. 
%but  semi-deterministic TWCs \cite{Chaaban:2017} may not satisfy the Shannon condition.
%Although the connection between Propositions~1 and~2 is not fully understood, the invariance of $\mathcal{H}(P_{X_2}\tilde{P}_{X_1|X_2}, P_{Y_j|X_1, X_2})$ in Proposition~2 in fact imposes a certain symmetry constraint on the channel marginal distribution $P_{Y_j|X_1, X_2}$.

% % % % % % % % % % % % % % % % % % % % % % % % % % % % % % % % % % % % % % % % % % % % % % % % %
\section{Conditions for the Tightness of Shannon's Inner and Outer Bounds}\label{subsec:newcondset}
In this section, we present four results regarding the tightness of Shannon's inner and outer bounds. 
We adopt the viewpoint that a two-way channel consists of two one-way channels with state. 
For example, the one-way channel from user~$1$ to user~$2$ is governed by the marginal distribution $P_{Y_2|X_1, X_2}$ (derived from the channel probability distribution $P_{Y_1, Y_2|X_1, X_2}$), where $X_1$ and $Y_2$ are respectively the input and the output of the channel with state $X_2$. 

Let $P_X$ and $P_{Y|X}$ be probability distributions on finite sets $\mathcal{X}$ and $\mathcal{Y}$. 
To simplify the presentation, we define 
\begin{IEEEeqnarray}{l}
\scalemath{0.87}{\mathcal{I}(P_{X}, P_{Y|X})=\sum\limits_{x, y}P_{X}(x)P_{Y|X}(y|x)\log%\nonumber\\ 
\frac{P_{Y|X}(y|x)}{\sum_{x'}P_{X}(x')P_{Y|X}(y|x')}},\nonumber\label{SMI}
\end{IEEEeqnarray}
which is the mutual information $I(X; Y)$ between input $X$ (governed by $P_X$) and corresponding output $Y$ of a channel with transition probability $P_{Y|X}$.
A useful fact is that $\mathcal{I}(\cdot, \cdot)$ is concave in the first argument when the second argument is fixed.  
Moreover, the conditional mutual information $I(X_1; Y_2|X_2=x_2)$ and $I(X_2; Y_1|X_1=x_1)$ can be expressed as $\mathcal{I}(P_{X_1|X_2=x_2}, P_{Y_2|X_1, X_2=x_2})$ and $\mathcal{I}(P_{X_2|X_1=x_1}, P_{Y_1|X_1=x_1, X_2})$, respectively. 

By viewing a TWC as two one-way channels with state, each of the following four theorems comprises two conditions, one for each direction of the two-way transmission. 
By symmetry, these theorems are also valid if the roles of users~1 and~2 are swapped. 
For simplicity, we will use $I^{(k)}(X_i; Y_j|X_j)$ and $H^{(k)}(Y_j|X_1, X_2)$ to denote the conditional mutual information and conditional entropy evaluated under input distribution $P^{(k)}_{X_1, X_2}$ for $i, j=1, 2$ with $i\neq j$. 
For $P^{(k)}_{X_1, X_2}=P^{(k)}_{X_j}P^{(k)}_{X_j|X_i}$ with $i\neq j$, the conditional entropy $H^{(k)}(Y_i|X_i)$ is evaluated under the marginal distribution $P^{(k)}_{Y_i|X_i}(y_i|x_i)=\sum_{x_j}P_{X_j|X_i}^{(k)}(x_j|x_i)P_{Y_i|X_j, X_i}(y_i|x_j, x_i)$. 
%\vspace{+0.1cm}

\begin{theorem}
\label{thm:23}
For a given memoryless TWC, if both of the following conditions are satisfied, then $\mathcal{C}_I=\mathcal{C}_O$: 
\begin{itemize}
\item[(i)] There exists $P^*_{X_1}\in\mathcal{P}(\mathcal{X}_1)$ such that for all $x_2\in\mathcal{X}_2$ we have $\argmax_{P_{X_1|X_2=x_2}} I(X_1; Y_2|X_2=x_2)=P^*_{X_1}$.  
%the optimal input distributions for the channel from user~1 to user~2 under different channel state are the same, i.e., 
\item[(ii)] $\mathcal{I}(P_{X_2}, P_{Y_1|X_1=x_1, X_2})$ does not depend on $x_1\in\mathcal{X}_1$ for any fixed $P_{X_2}\in\mathcal{P}(\mathcal{X}_2)$. 
%the mutual information of the channel from user~2 to user~1 under different channel state are identical when the same input distribution $P_{X_2}$ is used, i.e., 
\end{itemize}
\end{theorem}
\begin{IEEEproof}
For any $P^{(1)}_{X_1, X_2}=P^{(1)}_{X_2}P^{(1)}_{X_1|X_2}$, let $P^{(2)}_{X_1, X_2}=P^*_{X_1}P^{(1)}_{X_2}$, where $P^*_{X_1}$ is given by (i). 
In light of (i), we have
\begin{IEEEeqnarray}{l}
% & & \nonumber\\
I^{(1)}(X_1; Y_2|X_2)\nonumber \\
\ \ =\sum_{x_2}P^{(1)}_{X_2}(x_2)\cdot I^{(1)}(X_1; Y_2|X_2=x_2)\qquad\label{s3}\\
\ \ \le  \sum_{x_2}P^{(1)}_{X_2}(x_2)\cdot\scalemath{0.87}{\left[\max_{P_{X_1|X_2=x_2}}I(X_1; Y_2|X_2=x_2)\right]}\\
\ \ =  \sum_{x_2}P^{(1)}_{X_2}(x_2)\cdot \mathcal{I}(P^*_{X_1}, P_{Y_2|X_1, X_2=x_2})\\
\ \ = \sum_{x_2}P^{(1)}_{X_2}(x_2)\cdot I^{(2)}(X_1; Y_2|X_2=x_2)\\
\ \ = I^{(2)}(X_1; Y_2|X_2)\label{s4}.
\end{IEEEeqnarray}
Moreover, 
\begin{IEEEeqnarray}{l}
I^{(1)}(X_2; Y_1|X_1)\nonumber\\
\ \ = \sum_{x_1}P^{(1)}_{X_1}(x_1)\cdot I^{(1)}(X_2; Y_1|X_1=x_1)\label{cond3s1}\\
\ \ =\sum_{x_1}P^{(1)}_{X_1}(x_1)\cdot\mathcal{I}(P^{(1)}_{X_2|X_1=x_1}, P_{Y_1|X_1=x_1, X_2})\\
\ \ =\sum_{x_1}P^{(1)}_{X_1}(x_1)\cdot\mathcal{I}(P^{(1)}_{X_2|X_1=x_1}, P_{Y_1|X_1=x'_1, X_2})\label{cond3s2}\\
\ \ \le \scalemath{0.87}{\mathcal{I}\left(\sum_{x_1}P^{(1)}_{X_1}(x_1)P^{(1)}_{X_2|X_1}(x_2|x_1), P_{Y_1|X_1=x'_1, X_2}\right)}\label{cond3s3}\\
\ \ = \mathcal{I}(P^{(1)}_{X_2}, P_{Y_1|X_1=x'_1, X_2})\\
\ \ = \sum_{x'_1}{P}^*_{X_1}(x'_1)\cdot\mathcal{I}(P^{(1)}_{X_2}, P_{Y_1|X_1=x'_1, X_2})\label{cond3s5}\\
\ \ = I^{(2)}(X_2; Y_1|X_1)\label{cond3s6}, 
\end{IEEEeqnarray}
where (\ref{cond3s2}) holds by the invariance assumption in (ii), (\ref{cond3s3}) holds since the functional $\mathcal{I}(\cdot, \cdot)$ is concave in the first argument, and (\ref{cond3s5}) is obtained from the invariance assumption in (ii). 
Combining the above yields $\mathcal{R}(P^{(1)}_{X_1,X_2}, P_{Y_1,Y_2|X_1,X_2})\subseteq \mathcal{R}(P^*_{X_1}P^{(1)}_{X_2}, P_{Y_1,Y_2|X_1,X_2})$, which implies that $C_{O}\subseteq C_{I}$ and hence $C_{I}=C_{O}$.  
\end{IEEEproof}

\begin{theorem}
\label{thm:24}
For a given memoryless TWC, if for any $P_{X_1, X_2}=P_{X_2}P_{X_1|X_2}=P_{X_1}P_{X_2|X_1}$, both of the following conditions are satisfied, then $\mathcal{C}_I=\mathcal{C}_O$:
\begin{itemize}
\item[(i)] There exists $P^*_{X_1}\in\mathcal{P}(\mathcal{X}_1)$ such that for all $x_2\in\mathcal{X}_2$ we have $\argmax_{P_{X_1|X_2=x_2}} I(X_1; Y_2|X_2=x_2)=P^*_{X_1}$.   
\item[(ii)] $\mathcal{H}(P_{X_2}\tilde{P}_{X_1|X_2}, P_{Y_1|X_1, X_2})$ does not depend on $\tilde{P}_{X_1|X_2}$ given $P_{X_2}$ and $P_{Y_1|X_1, X_2}$, and the common maximizer $P^*_{X_1}$ in (i) also satisfies  
$\scalemath{0.98}{\mathcal{\bar{H}_{\perp}}(P^*_{X_1}, P_{X_2}, P_{Y_1|X_1, X_2})\ge\bar{\mathcal{H}}(P_{X_1}, P_{X_2|X_1}, P_{Y_1|X_1, X_2})}.$
\end{itemize}
\end{theorem}
\begin{IEEEproof}
Given any $P^{(1)}_{X_1, X_2}=P^{(1)}_{X_2}P^{(1)}_{X_1|X_2}$, let $P^{(2)}_{X_1, X_2}=P^*_{X_1}P^{(1)}_{X_2}$. 
By the same argument as in (\ref{s3})-(\ref{s4}), we obtain via (i) that $I^{(1)}(X_1; Y_2|X_2)\le I^{(2)}(X_1; Y_2|X_2)$. 
Moreover,  
\begin{IEEEeqnarray}{l}
I^{(1)}(X_2; Y_1|X_1)\nonumber\\
\ \ =H^{(1)}(Y_1|X_1)-H^{(1)}(Y_1|X_1, X_2)\nonumber\\
\ \ =\scalemath{0.9}{\mathcal{\bar{H}}(P^{(1)}_{X_1}, P^{(1)}_{X_2|X_1}, P_{Y_1|X_1, X_2})-\mathcal{H}(P^{(1)}_{X_2}P^{(1)}_{X_1|X_2}, P_{Y_1|X_1, X_2})}\label{thm23s3}\IEEEeqnarraynumspace\\
\ \ \le \scalemath{0.9}{\mathcal{\bar{H}_{\perp}}(P^*_{X_1}, P^{(1)}_{X_2}, P_{Y_1|X_1, X_2})-\mathcal{H}(P^{(1)}_{X_2}P^*_{X_1}, P_{Y_1|X_1, X_2})}\label{thm23s4}\\
\ \ = H^{(2)}(Y_1|X_1)-H^{(2)}(Y_1|X_1, X_2)\label{thm23s5}\\
\ \ = I^{(2)}(X_2; Y_1|X_1),\nonumber 
\end{IEEEeqnarray}
where (\ref{thm23s3}) and (\ref{thm23s5}) follow from the definitions in Section~II and (\ref{thm23s4}) is due to condition (ii). 
Consequently, $\mathcal{R}(P^{(1)}_{X_1,X_2}, P_{Y_1,Y_2|X_1,X_2})\subseteq \mathcal{R}(P^*_{X_1}P^{(1)}_{X_2}, P_{Y_1,Y_2|X_1,X_2})$, and hence $\mathcal{C}_O\subseteq\mathcal{C}_I$, so that $\mathcal{C}_I=\mathcal{C}_O$. 
\end{IEEEproof}

\begin{theorem}
\label{thm:33}
For a given memoryless TWC, if both of the following conditions are satisfied, then $\mathcal{C}_I=\mathcal{C}_O$: 
\begin{itemize}
\item[(i)] $\mathcal{I}(P_{X_1}, P_{Y_2|X_1, X_2=x_2})$ does not depend on $x_2\in\mathcal{X}_2$ for any fixed $P_{X_1}\in\mathcal{P}(\mathcal{X}_1)$.
\item[(ii)] $\mathcal{I}(P_{X_2}, P_{Y_1|X_1=x_1, X_2})$ does not depend on $x_1\in\mathcal{X}_1$ for any fixed $P_{X_2}\in\mathcal{P}(\mathcal{X}_2)$.
\end{itemize}
\end{theorem}
\begin{IEEEproof}
From conditions (i) and (ii), we know that $\max_{P_{X_1|X_2=x_2}} I(X_1; Y_2|X_2=x_2)$ has a common maximizer $P^*_{X_1}$ for all $x_2\in\mathcal{X}_2$ and $\max_{P_{X_2|X_1=x_1}} I(X_2; Y_1|X_1=x_1)$ has a common maximizer $P^*_{X_2}$ for all $x_1\in\mathcal{X}_1$.  
For any $P^{(1)}_{X_1, X_2}=P^{(1)}_{X_1}P^{(1)}_{X_2|X_1}$, let $P^{(2)}_{X_1, X_2}=P^*_{X_1}P^*_{X_2}$. 
By the same argument as in (\ref{s3})-(\ref{s4}), we conclude that $I^{(1)}(X_1; Y_2|X_2)\le I^{(2)}(X_1; Y_2|X_2)$ and $I^{(1)}(X_2; Y_1|X_1)\le \allowbreak I^{(2)}(X_2; Y_1|X_1)$. 
Thus, $\mathcal{R}(P^{(1)}_{X_1,X_2}, P_{Y_1,Y_2|X_1,X_2})\subseteq \mathcal{R}(P^*_{X_1}P^*_{X_2}, P_{Y_1,Y_2|X_1,X_2})$, which yields $\mathcal{C}_I=\mathcal{C}_O$. 
\end{IEEEproof}

Similar to the CVA condition, complex computations are often inevitable for checking the above conditions.  
We next present a useful condition which needs little computational effort. 
Let $[P_{Y_2|X_1, X_2}(\cdot|\cdot, x_2)]$ (resp. $[P_{Y_1|X_1, X_2}(\cdot|x_1, \cdot)]$) denote the marginal transition probability matrix obtained from $P_{Y_1, Y_2|X_1, X_2=x_2}$ (resp. $P_{Y_1, Y_2|X_1=x_1, X_2}$), whose columns and rows are indexed according to a fixed order on the symbols in $\mathcal{Y}_2$ and $\mathcal{X}_1$ (resp. $\mathcal{Y}_1$ and $\mathcal{X}_2$). 
\begin{theorem}
\label{thm:77}
For a given memoryless TWC, if both of the following conditions are satisfied, then $\mathcal{C}_I=\mathcal{C}_O$: 
\begin{itemize}
\item[(i)] The matrices $[P_{Y_2|X_1, X_2}(\cdot|\cdot, x_2)]$, $x_2\in\mathcal{X}_2$, are column permutations of each other. 
\item[(ii)] The matrices $[P_{Y_1|X_1, X_2}(\cdot|x_1, \cdot)]$, $x_1\in\mathcal{X}_1$, are column permutations of each other.
\end{itemize}
\end{theorem}
Since the proof is similar to the second part of the proof of Theorem~\ref{thm:23toSC} in the next section, the details are omitted. 

\vspace{-0.1cm}
\section{Discussion and Examples}
\subsection{Comparison with Other Conditions}
As already noted, the relationship between Propositions~\ref{thm:SC} and \ref{thm:CVA} is unclear as examples that satisfy the Shannon condition but not the CVA condition seem hard to construct. In this section, we show that Theorems~\ref{thm:23} and~\ref{thm:24} in fact generalize the Shannon and CVA results, respectively. 
To see this, it suffices to show that the Shannon and CVA conditions imply the conditions in Theorems~\ref{thm:23} and \ref{thm:24}, respectively. 

\begin{theorem}
A TWC satisfying Shannon's symmetry condition in Proposition~1 must satisfy the conditions in Theorem~\ref{thm:23}. 
\label{thm:23toSC}
\end{theorem}
\vspace{-0.4cm}
\begin{IEEEproof}
For a TWC satisfying the condition of Proposition~\ref{thm:SC}, the optimal input probability distribution that achieves capacity is of the form $P_{X_1, X_2}=P_{X_2}/|\mathcal{X}_1|$ for some $P_{X_2}\in\mathcal{P}(\mathcal{X}_2)$ \cite{Shannon:1961}. 
This result implies that condition (i) of Theorem~\ref{thm:23} is satisfied because a common maximizer exists for all $x_2\in\mathcal{X}$ and is given by $P^*_{X_1}(x_1)=1/|\mathcal{X}_1|$. 
To prove that condition (ii) is also satisfied, we consider the two (marginal) matrices $[P_{Y_1|X_1, X_2}(\cdot|x'_1, \cdot)]$ and $[P_{Y_1|X_1, X_2}(\cdot|x''_1, \cdot)]$ for some fixed $x'_1, x''_1\in\mathcal{X}_1$ and show that these matrices are column permutations of each other and hence $\mathcal{I}(P_{X_2}, P_{Y_1|X_1=x'_1, X_2})=\mathcal{I}(P_{X_2}, P_{Y_1|X_1=x''_1, X_2})$. 
The former claim is true because
\begin{IEEEeqnarray}{l}
P_{Y_1|X_1,X_2}(y_1|x'_1,x_2)\quad \nonumber\\
\ \ = P_{Y_1|X_1,X_2}(\pi_1^{\mathcal{Y}_1}[x_1',x_1''](y_1)|\tau^{\mathcal{X}_1}_{x_1',x_1''}(x'_1),x_2)\label{eq:com11}\\
\ \  = P_{Y_1|X_1,X_2}(\pi_1^{\mathcal{Y}_1}[x_1',x_1''](y_1)|x''_1,x_2),\label{eq:com12}
\end{IEEEeqnarray}
where \eqref{eq:com11} is obtained by marginalizing $Y_2$ on both sides of (\ref{SCcond}) and \eqref{eq:com12} follows from the definition of transposition. 
The second claim can be verified by a direct computation on $\mathcal{I}(P_{X_2}, P_{Y_1|X_1=x_1, X_2})$ with the above result straightforwardly, and hence the details are omitted.  
%\begin{IEEEeqnarray}{rCl}
%& &\ \mathcal{I}(P_{X_2}, P_{Y_1|X_1=x'_1, X_2})\nonumber\\
%&=& \scalemath{0.9}{\sum_{x_2, y_1}P_{X_2}(x_2)P_{Y_1|X_1, X_2}(y_1|x'_1, x_2)}\nonumber\\ [-0.4cm]
%& &\qquad\qquad\qquad \scalemath{0.9}{\log\frac{P_{Y_1|X_1, X_2}(y_1|x'_1, x_2)}{\sum_{x'_2}P_{X_2}(x'_2)P_{Y_1|X_1, X_2}(y_1|x'_1, x'_2)}}\nonumber\\
%&=& \scalemath{0.9}{\sum_{x_2, y_1}P_{X_2}(x_2)P_{Y_1|X_1, X_2}(\pi^{\mathcal{Y}_1}[x'_1, x''_1](y_1)|x''_1, x_2)}\nonumber\\[-0.4cm]
%& &\qquad\scalemath{0.9}{\log\frac{P_{Y_1|X_1, X_2}(\pi^{\mathcal{Y}_1}[x'_1, x''_1](y_1)|x''_1, x_2)}{\sum_{x'_2}P(x'_2)P_{Y_1|X_1, X_2}(\pi^{\mathcal{Y}_1}[x'_1, x''_1](y_1)|x''_1, x_2)}}\label{C7toC3s1}\IEEEeqnarraynumspace\\
%&=& \scalemath{0.9}{\sum_{x_2, y'_1}P_{X_2}(x_2)P_{Y_1|X_1, X_2}(y'_1|x''_1, x_2)}\nonumber\\[-0.4cm]
%& &\qquad\qquad\qquad\scalemath{0.9}{\log\frac{P_{Y_1|X_1, X_2}(y'_1|x''_1, x_2)}{\sum_{x_2^{'}}P_{X_2}(x'_2)P_{Y_1|X_1, X_2}(y'_1|x''_1, x_2)}}\label{C7toC3s2}\nonumber\\[+0.1cm]
%&=&\ \mathcal{I}(P_{X_2}, P_{Y_1|X_1=x''_1, X_2})\nonumber\label{C7toC3s3},
%\end{IEEEeqnarray}
%where (\ref{C7toC3s1}) holds by the first claim. 
\end{IEEEproof}

\begin{remark}
Example~1 in the next subsection demonstrates that a TWC that satisfies the conditions in Theorem~\ref{thm:23} may not satisfy Shannon's symmetry condition in Proposition~\ref{thm:SC} since the common maximizer is not necessarily the uniform input distribution. Hence, Theorem~\ref{thm:23} is a more general result than Proposition~\ref{thm:SC}. %An example illustrating this observation is given later. 
%Furthermore, based on the second part of the proof for Theorem~\ref{thm:23toSC}, it can be shown that if a TWC satisfies Shannon's symmetry condition for both users, then it must satisfy the conditions of Theorems~\ref{thm:33} and~\ref{thm:77}. 
\end{remark}

\begin{theorem}
A TWC satisfying the CVA condition in Proposition~2 must satisfy the conditions in Theorem~\ref{thm:24}. 
\label{thm:CVAto24}
\end{theorem}
\begin{IEEEproof}
Suppose that the condition of Proposition~\ref{thm:CVA} is satisfied.  
To prove the theorem, we first claim that for $j=1, 2$, $H(Y_j|X_1=x'_1, X_2=x'_2)=H(Y_j|X_1=x''_1, X_2=x'_2)$ for all $x'_1, x''_1\in\mathcal{X}_1$ and $x'_2\in\mathcal{X}_2$. 
Given arbitrary pairs $(x'_1, x'_2)$ and $(x''_1, x'_2)$ with $x'_1\neq x''_1$, consider the two probability distributions 
\begin{equation}
P^{(1)}_{X_1, X_2}(a, b)=\left\{
\begin{array}{ll}
1,&\ \text{if}\ a=x'_1\ \text{and}\ b=x'_2,\\
0,&\ \text{otherwise,}
\end{array}
\right.\nonumber
\end{equation}
and
\begin{equation}
P^{(2)}_{X_1, X_2}(a, b)=\left\{
\begin{array}{ll}
1,&\ \text{if}\ a=x''_1\ \text{and}\ b=x'_2,\\
0,&\ \text{otherwise.}
\end{array}
\right.\nonumber
\end{equation}
Noting that $P^{(1)}_{X_2}=P^{(2)}_{X_2}$, we have
\begin{IEEEeqnarray}{rCl}
H(\scalemath{0.9}{Y_j|X_1=x'_1, X_2=x'_2})&=&H^{(1)}(\scalemath{0.9}{Y_j|X_1, X_2})\IEEEeqnarraynumspace\label{thm5s1}\\
&=&\mathcal{H}(\scalemath{0.9}{P^{(1)}_{X_2}P^{(1)}_{X_1|X_2}, P_{Y_j|X_1, X_2}})\nonumber\label{thm5s2}\\
&=&\mathcal{H}(\scalemath{0.9}{P^{(1)}_{X_2}P^{(2)}_{X_1|X_2}, P_{Y_j|X_1, X_2}})\IEEEeqnarraynumspace\label{thm5s3}\\
&=&H^{(2)}(\scalemath{0.9}{Y_j|X_1, X_2})\label{thm5s4}\IEEEeqnarraynumspace\\
&=&H(\scalemath{0.9}{Y_j|X_1=x''_1, X_2=x'_2}),\label{thm5s5}\IEEEeqnarraynumspace
\end{IEEEeqnarray}
where (\ref{thm5s1}) and (\ref{thm5s5}) are due to the definitions of $P^{(1)}_{X_1, X_2}$ and $P^{(2)}_{X_1, X_2}$, respectively, (\ref{thm5s3}) follows from the CVA condition, and (\ref{thm5s4}) holds since $P^{(1)}_{X_2}=P^{(2)}_{X_2}$. 
The claim is proved.
Since $H(Y_j|X_1, X_2=x_2)=\sum_{x_1}P_{X_1|X_2}(x_1|x_2)H(Y_j|X_1=x_1, X_2=x_2)$ and $H(Y_j|X_1=x_1, X_2=x_2)$ does not depend on $x_1\in\mathcal{X}_1$ for fixed $x_2\in\mathcal{X}_2$, $H(Y_j|X_1, X_2=x_2)$ does not depend on $P_{X_1|X_2=x_2}$. 

Next, we show that condition (i) of Theorem~\ref{thm:24} holds by constructing the common maximizer from the CVA condition. 
For each $x_2\in\mathcal{X}_2$, let $P^*_{X_1|X_2=x_2}=\argmax_{P_{X_1|X_2=x_2}}I(X_1; Y_2|X_2=x_2)=\argmax_{P_{X_1|X_2=x_2}} [H(Y_2|X_2=x_2)-H(Y_2|X_1, X_2=x_2)]$ and define $P^{(1)}_{X_1, X_2}=P^{(1)}_{X_2}P^*_{X_1|X_2}$ for some $P^{(1)}_{X_2}\in\mathcal{P}(\mathcal{X}_2)$. 
Since $H(Y_j|X_1, X_2=x_2)$ does not depend on $P_{X_1|X_2=x_2}$, $P^*_{X_1|X_2=x_2}$ is in fact a maximizer for $H(Y_2|X_2=x_2)$. 
Note that the maximizer $P^*_{X_1|X_2=x_2}$ may not be unique, but any choice works for our purposes.
Now for $P^{(1)}_{X_1, X_2}$, by the CVA condition, there exists $\tilde{P}_{X_1}\in\mathcal{P}(\mathcal{X}_1)$ such that 
\[
\bar{\mathcal{H}}(P^{(1)}_{X_2}, P^*_{X_1|X_2}, P_{Y_2|X_1, X_2})\le \mathcal{\bar{H}_{\perp}}(P^{(1)}_{X_2}, \tilde{P}_{X_1}, P_{Y_2|X_1, X_2}). 
\]
Set $P^{(2)}_{X_1, X_2}=\tilde{P}_{X_1}P^{(1)}_{X_2}$. Since $P^*_{X_1|X_2=x_2}$ is the maximizer for $H(Y_2|X_2=x_2)$, we have
\begin{IEEEeqnarray}{l}
\mathcal{\bar{H}}(P^{(1)}_{X_2}, P^*_{X_1|X_2}, P_{Y_2|X_1, X_2})\qquad \nonumber\\
\ \ = H^{(1)}(Y_2|X_2)\nonumber\\
\ \ =\sum_{x_2}P^{(1)}_{X_2}(x_2)\cdot H^{(1)}(Y_2|X_2=x_2)\nonumber\\
\ \ =\sum_{x_2}P^{(1)}_{X_2}(x_2)\cdot\scalemath{0.9}{\left[\max_{P_{X_1|X_2=x_2}}H(Y_2|X_2=x_2)\right]}\nonumber\\
\ \ \ge  \sum_{x_2}P^{(1)}_{X_2}(x_2)\cdot H^{(2)}(Y_2|X_2=x_2)\nonumber\label{thm6:s4}\\
\ \ = H^{(2)}(Y_2|X_2)\nonumber\\
\ \ = \mathcal{\bar{H}}_{\perp}(P^{(1)}_{X_2}, \tilde{P}_{X_1}, P_{Y_2|X_1, X_2}).\nonumber 
\end{IEEEeqnarray}
Thus, $\scalemath{0.95}{\bar{\mathcal{H}}(\scalemath{1}{P^{(1)}_{X_2}, P^*_{X_1|X_2}, P_{Y_2|X_1, X_2}})=\mathcal{\bar{H}_{\perp}}(\scalemath{1}{P^{(1)}_{X_2}, \tilde{P}_{X_1}, P_{Y_2|X_1, X_2}})}$, i.e., 
\begin{IEEEeqnarray}{l}
\scalemath{0.87}{\sum\limits_{x_2}P^{(1)}_{X_2}(x_2)\cdot H^{(1)}(Y_2|X_2=x_2)=\sum\limits_{x_2}P^{(1)}_{X_2}(x_2)\cdot H^{(2)}(Y_2|X_2=x_2)}.\nonumber
\end{IEEEeqnarray}
Since $H^{(2)}(Y_2|X_2=x_2)\le H^{(1)}(Y_2|X_2=x_2)$ for each $x_2\in\mathcal{X}_2$, we obtain $H^{(1)}(Y_2|X_2=x_2)=H^{(2)}(Y_2|X_2=x_2)$, i.e., $\tilde{P}_{X_1}$ achieves the same value of $H(Y_2|X_2=x_2)$ as $P^*_{X_1|X_2=x_2}$ for all $x_2\in\mathcal{X}_2$. Consequently, $\tilde{P}_{X_1}$ is a common maximizer and thus condition (i) of Theorem~\ref{thm:24} is satisfied.
Moreover, since the common maximizer $\tilde{P}_{X_1}$ is provided by the CVA condition, condition (ii) of Theorem~\ref{thm:24} automatically holds.
\end{IEEEproof}

\begin{remark}
Example~1 below shows that a TWC that satisfies the conditions in Theorem~\ref{thm:24} does not necessarily satisfy the condition in Proposition~\ref{thm:CVA} because our conditions allow $\mathcal{H}(P_{X_2}\tilde{P}_{X_1|X_2}, P_{Y_2|X_1, X_2})$ to depend on $\tilde{P}_{X_1|X_2}$ for given $P_{X_2}$. Hence, Theorem~\ref{thm:24} is more general than Proposition~\ref{thm:CVA}. 
\end{remark}

\subsection{Examples}
We next illustrate the effectiveness of our conditions via two examples in which $\mathcal{X}_1=\mathcal{X}_2=\mathcal{Y}_1=\mathcal{Y}_2=\{0, 1\}$. 
The TWC in Example~1 satisfies the conditions of Theorems~\ref{thm:23}-\ref{thm:77} and the capacity region is rectangular.
The TWC in Example~2 satisfies the conditions of Theorem~\ref{thm:23} and~\ref{thm:24} and has a non-rectangular capacity region. 
However, neither of the constructed TWCs satisfy the Shannon or the CVA conditions. 

\begin{example}
Consider the TWC with %\vspace{-0.3cm}
\begin{align*}
\scalemath{0.87}{
[P_{Y_1,Y_2|X_1,X_2}(\cdot, \cdot|\cdot, \cdot)] = 
 \bordermatrix{
& 00& 01 & 10 & 11\cr
00& 0.783 & 0.087 & 0.117 & 0.013\cr
01& 0.0417 & 0.3753 & 0.0583 & 0.5247\cr
10& 0.261 & 0.609 & 0.039 & 0.091 \cr
11& 0.2919 & 0.1251 & 0.4081 & 0.1749}.}
\end{align*}
The corresponding one-way channel marginal distributions are given by
\begin{align*}
\scalemath{0.83}{
{[P_{Y_2|X_1,X_2}(\cdot|\cdot,0)]= %\atop 
\begin{pmatrix}
0.9 & 0.1\\
0.3 & 0.7
\end{pmatrix},}}\hspace{+0.1cm}
\scalemath{0.83}{{[P_{Y_1|X_1,X_2}(\cdot|0,\cdot)]= %\atop 
\begin{pmatrix}
0.87 & 0.13\\
0.417 & 0.583
\end{pmatrix},}}\\[+0.1cm]
\scalemath{0.83}{{[P_{Y_2|X_1,X_2}(\cdot|\cdot,1)]= %\atop 
 \begin{pmatrix}
0.1 & 0.9\\
0.7 &  0.3
\end{pmatrix},}}\hspace{+0.1cm}
\scalemath{0.83}{
{[P_{Y_1|X_1,X_2}(\cdot|1,\cdot)]= %\atop 
 \begin{pmatrix}
0.87 & 0.13\\
0.417 &  0.583
\end{pmatrix}.}}
\end{align*}
For this TWC, Shannon's symmetry condition in Proposition~\ref{thm:SC} does not hold since there are no permutations on $\mathcal{Y}_1$ and $\mathcal{Y}_2$ which can result in (\ref{SCcond}). Furthermore, since $H(Y_2|X_1=0, X_2=0)=H_{\text{b}}(0.1)$ and $H(Y_2|X_1=1, X_2=0)=H_{\text{b}}(0.3)$, where $H_{\text{b}}(\cdot)$ denotes the binary entropy function, $\mathcal{H}(P_{X_2}\tilde{P}_{X_1|X_2}, P_{Y_2|X_1, X_2})$ depends on $\tilde{P}_{X_1|X_2}$ for given $P_{X_2}$. 
Thus, the CVA condition in Proposition~\ref{thm:CVA} does not hold, either. 

However by Theorem~\ref{thm:77}, Shannon's inner and outer bounds coincide since $[P_{Y_2|X_1,X_2}(\cdot|\cdot,0)]$ (resp. $[P_{Y_1|X_1,X_2}(\cdot|0, \cdot)]$) can be obtained by permuting the columns of $[P_{Y_2|X_1,X_2}(\cdot|\cdot,1)]$ (resp. $[P_{Y_1|X_1,X_2}(\cdot|1, \cdot)]$). 
Since the conditions in Theorem~\ref{thm:77} imply the conditions in Theorem~\ref{thm:33} and the conditions in Theorem~\ref{thm:33} further imply the conditions in Theorem~\ref{thm:23}, the conditions of Theorems~\ref{thm:23} and~\ref{thm:33} are also satisfied.  
Moreover, the optimal input distribution for this TWC can be obtained by searching for the common maximizer for each of the two one-way channels via the Blahut-Arimoto algorithm yielding $P^*_{X_1}(0)=P^*_{X_2}(0)=0.471$. Thus, the capacity region is achieved by the input distribution $P^*_{X_1, X_2}=P^*_{X_1}P^*_{X_2}$, i.e., $\mathcal{C}=\{(R_1, R_2): 0\le R_1\le 0.2967, 0\le R_2\le 0.1715\}$. 

Finally, we note that this TWC also satisfies the conditions of Theorem~\ref{thm:24}, in which the first condition is already implied by the conditions of Theorem~\ref{thm:23}.  
To verify the second condition, we consider 
\begin{align*}
\scalemath{0.82}{{[P_{Y_1|X_1,X_2}(\cdot|\cdot,0)]= %\atop 
\begin{pmatrix}
0.87 & 0.13\\
0.87 & 0.13
\end{pmatrix},}}\hspace{+0.1cm}
\scalemath{0.82}{{[P_{Y_1|X_1,X_2}(\cdot|\cdot,1)]= %\atop 
 \begin{pmatrix}
0.417 & 0.583\\
0.417 &  0.583
\end{pmatrix}.}}
\end{align*}
Here, for all $x_1\in\{0, 1\}$, $H(Y_1|X_1=x_1, X_2=0)=H_{\text{b}}(0.13)$ and $H(Y_1|X_1=x_1, X_2=1)=H_{\text{b}}(0.417)$. 
Thus, $\mathcal{H}(P_{X_2}\tilde{P}_{X_1|X_2}, P_{Y_1|X_1, X_2})$ does not depend on $\tilde{P}_{X_1|X_2}$ given $P_{X_2}$. 
Together with the substitutions $P_{X_1,X_2}^{(1)}=P_{X_1, X_2}$ and $P_{X_1, X_2}^{(2)}=P^*_{X_1}P_{X_2}$ into (\ref{cond3s1})-(\ref{cond3s6}), we then obtain that $\mathcal{\bar{H}_{\perp}}(P^*_{X_1}, P_{X_2}, P_{Y_1|X_1, X_2})\ge\allowbreak\bar{\mathcal{H}}(P_{X_1}, P_{X_2|X_1}, P_{Y_1|X_1, X_2})$. Therefore, the second condition of Theorem~\ref{thm:24} holds. 
\end{example}
 %because $\mathcal{H}(P_{X_2}\tilde{P}_{X_1|X_2}, P_{Y_1|X_1, X_2})$ does not depend on $\tilde{P}_{X_1|X_2}$

\begin{figure}[tb]
\begin{centering}
\includegraphics[scale=0.293, draft=false]{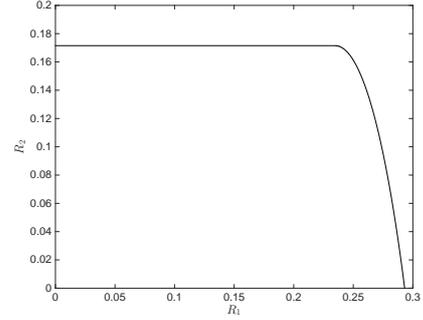}
\vspace{-0.1cm}
\caption{The capacity region of the TWC in Example 2.\label{fig:capacityExm}}
\end{centering}
\vspace{-0.35cm}
\end{figure}

\begin{example}
Consider the TWC with 
\begin{align*}
\scalemath{0.87}{[P_{Y_1,Y_2|X_1,X_2}] = 
 \bordermatrix{
& 00& 01 & 10 & 11\cr
00& 0.783 & 0.087 & 0.117 & 0.013\cr
01& 0.36279 & 0.05421 & 0.50721 & 0.07579\cr
10& 0.261 & 0.609 & 0.039 & 0.091 \cr
11& 0.173889 & 0.243111 & 0.243111 & 0.339889},}
\end{align*}
where two one-way channel marginal distributions are
\[
\scalemath{0.83}{{[P_{Y_2|X_1,X_2}(\cdot|\cdot,0)]= %\atop 
\begin{pmatrix}
0.9 & 0.1\\
0.3 & 0.7
\end{pmatrix},}}\hspace{+0.1cm}
\scalemath{0.83}{{[P_{Y_2|X_1,X_2}(\cdot|\cdot,1)]= %\atop 
 \begin{pmatrix}
0.87 & 0.13\\
0.417 &  0.583
\end{pmatrix},}}\\%[+0.1cm]
%\scalemath{0.83}{{[P_{Y_1|X_1,X_2}(\cdot|0,\cdot)]= %\atop 
%\begin{pmatrix}
%0.87 & 0.13\\
%0.417 & 0.583
%\end{pmatrix},}}\hspace{+0.1cm}
%\scalemath{0.83}{{[P_{Y_1|X_1,X_2}(\cdot|1,\cdot)]= %\atop 
% \begin{pmatrix}
%0.87 & 0.13\\
%0.417 &  0.583
%\end{pmatrix}.}}
\]
$[P_{Y_1|X_1,X_2}(\cdot|0,\cdot)]=[P_{Y_1|X_1,X_2}(\cdot|1,\cdot)]=[P_{Y_2|X_1,X_2}(\cdot|\cdot,1)]$.
Using the same arguments as in Example~1, one can easily see that this TWC satisfies neither the Shannon nor the CVA conditions. 
However, it satisfies the conditions in Theorem~\ref{thm:23} since a common maximizer exists for the one-way channel from users~1 to~2, i.e., $P^*_{X_1}(0)=0.471$, and condition~(ii) trivially holds. 
To verify that this channel also satisfies the conditions in Theorem~\ref{thm:24}, the same argument as in the previous example is used.  
Finally, by considering all input distributions of the form $P_{X_1, X_2}=P^*_{X_1}P_{X_2}$, the capacity region of this channel is determined as shown in Fig.~\ref{fig:capacityExm}. 
\end{example}

\vspace{-0.1cm}
\section{Conclusions}
In this paper, four conditions on the coincidence of Shannon's capacity inner and outer bounds were derived. 
These invariance conditions were shown to generalize existing results, thus enlarging the class of TWCs whose capacity region can be exactly determined. 
Numerical examples illustrate the applications of the new conditions in situations where prior results do not apply.

\end{document}